# Microwave impedance of a dc-biased Josephson Fluxonic Diode in the presence of magnetic field and rf drive

*Hamed Mehrara, Alireza Erfanian, Farshid Raissi*

The dependence of microwave impedance of a dc-biased Josephson Fluxonic Diode (JFD) under application of both dc magnetic field and rf excitation is calculated with a variety of conditions. For finite length of a JFD excited by a very low microwave excitation below its plasma frequency, applied dc magnetic field increases the rate of Vortex and Anti-Vortex (VAV) pair generation which fine-tunes the microwave resistance up to several factors more than its zero field microwave resistance ($R_0$). Under this circumstance, adding a dc bias for moving VAVs causes oscillation-like features in microwave impedance of JFD either in forward or reverse bias. As a result, the microwave resistance increases up to $30R_0$ in the forward bias despite the fact that damping parameter ($\beta$) can limit this increase. On the other hand, sharp phase slips are seen in reverse bias mode on the reactance of overdamped JFD while increasing the frequency or amplitude of microwave excitation leads to unprecedented effects of resistance which is described.

## I. INTRODUCTION

The response of Josephson junctions (JJs) to rf excitation has been the subject of an enormous amount of research over the last decades. Shapiro steps [1], I-V characteristics [2], microwave impedance [3] and chaos [4, 5] are several cases in this field which have been extensively studied both theoretically and experimentally for these family of superconducting devices similar to single crystals [6], weak links [7, 8], SIS [9] and long Josephson Junctions (LJJs) [10]. The microwave impedance of these intrinsic and extrinsic devices has shown a remarkable nonlinear dependence to the rf drive which can be described in terms of Josephson-like response. Among them, extrinsic devices allow improved microwave sensitivity and dynamic range because of the lower critical current densities and response times; however at the present time, different variation to extrinsic Josephson junctions with microwave application potential are Superconducting Fluxonic Electronics (SFE) which continue to be of considerable interest [11]. Such benefits arise from the dynamic of fluxons (which is also called “vortex” or “soliton”) and rf characteristics of corresponding devices. Much of this interest stems from the desire to use high-temperature superconductors (HTS's) components in passive microwave devices such as oscillators, mixer, filters and impedance matching lines for wireless communication

The motivation of this study is to investigate microwave properties for a Josephson fluxonic diode (JFD) with its unique vortex dynamics at rectifying the supercurrent in SFE. Also, this dynamics have potential application in circuitry of flux qubits readout, DC/SFQ and SFQ/DC convertors and high frequency rapid single flux quantum (RSFQ) electronics [12]. Josephson fluxonic diode is dual of semiconductor p-n junction in which fluxon motion can be generated by a uniform dc biasing with an external spatially reversing magnetic field. It provides an asymmetric IVC where the forward bias corresponds to a voltage state and reverse bias to a short circuit that could be in principle useful for detecting, mixing and switching microwave signals [13].

In this paper, we report systematic study which attempts to define microwave impedance of a JFD under important class of experiments which is separate and simultaneous dc and rf excitations field with biasing current. At this time, it is important to consider both symmetric and anti-symmetric microwave excitation where the field direction either matches or is in the opposite direction to the external dc magnetic field applied to the sides of the JFD. It is also essential to consider the impedance characteristics of this device versus frequency and amplitude of excitation with damping parameter relation in a simulated model. So, Sec. II contains theoretical background around Josephson fluxonic diode and in Sec. III, the implications of dc magnetics field, bias current, amplitude and frequency of microwave excitation are described step by step. Finally, the results are summarized in Sec. IV.

## II. BACKGROUND

JFD is a long Josephson junction to which a spatially reversing magnetic field is applied to generates soliton carriers in it [12]. Fig. 1 demonstrates the configuration of a JFD. Vortices fill one half and anti-vortices fill the other half of JFD as magnetic field carriers in its Josephson junction medium. Generated VAV pairs are uniformly distributed in the junction and the distance at the center of the JFD, in which the polarity of the magnetic field changes, is termed "transition region". The dc magnetic field which is applied by "control" line is perpendicular to the surface of the JFD (black solid line in Fig. 1) and there is a "bias" current in which uniformly passes through the JFD. It holds an asymmetric I-V curve based on bias polarity. In forward bias, vortices and anti-vortices are pushed toward the transition region and injected to the other side or annihilated at transition region based on damping properties of JFD; therefore creates a voltage across the JFD. In reverse bias, they are only pushed toward the ends and in steady state no movement of fluxons can be maintained; therefore, the device exhibits a short circuit feature. However, if the magnetic field is large enough, at a certain bias, a "breakdown" type behavior happens and VAV pairs are created at the center of JFD in transition region, result in a sudden voltage drop.

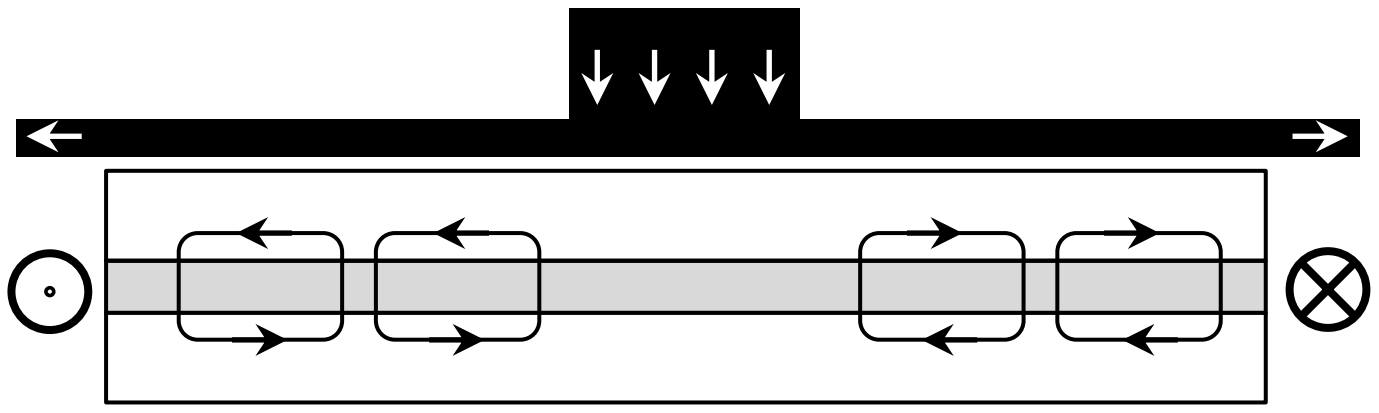

Fig. 1. Configuration for a JFD: a separate control line running above the base electrode of a long Josephson junction creates a spatially reversing magnetic field inside it which generate vortices and anti-vortices in each side (dimensions are not to scale)

JFD's model consists of series connected Josephson junctions similar to a LJJ [14]. For a Josephson junction, total current $I$, flowing through the circuit has contributions from supercurrent $I_S$, normal current $I_N$ and displacement current $I_D$. While the normal current is caused by the flow of normal electrons in the barrier of Josephson junction, the displacement current is due to the time varying electric field across it. Therefore, for the total current $I_T$ passes through a JFD one can write:

$$I_T(t) = \sum_{x=1}^{N} (I_s(x,t) + I_N(x,t) + I_D(x,t)) \tag{1}$$

where $N$ is total number of ideal Josephson junction along length of JFD ($L$). Usually a Josephson junction as the unit cell of JFD models by ideal junction shunted with a resistance $R$ and capacitance $C$ to form a parallel circuit [14]. So one can easily rewrite Eq. 1 by means of stated parameters as below:

$$I_T(t) = \sum_{x=1}^{N} [I_c \sin(\varphi(x,t)) + \frac{\Phi_0}{2\pi R}\frac{d\varphi(x,t)}{dt} + \frac{\Phi_0 C}{2\pi}\frac{d^2\varphi(x,t)}{dt^2}] \tag{2}$$

Right-hand side of Eq. 2 represents components of the total current flowing through the JFD where $\varphi(x,t)$ and $I_c$ are the time varying phase and the critical current of each Josephson junction along the length of JFD, respectively. $\Phi_0$ is the unit of a flux quantum and here plasma frequency of a JFD, $\omega_p$, is defined by $\sqrt{(2\pi I_{cT}/\Phi_0 C_T)}$, where $I_{cT}$ is the sum of critical current over the entire length of JFD and $C_T$ is its total capacitance.

When a JFD is subjected to a pure microwave signal, the modulated rf current can represent as $I_{rf} \sin(\omega t)$ which can be applied symmetrically or asymmetrically to the both sides of JFD. It should be noted that the microwave drive can induce as a surface current either by the resonant cavity microwave magnetic field parallel to the direction of the JFD, or by the electric field coupled to the microwave source through capacitance coupling such as in strip line resonators. So $I_{rf}\,sin(\omega t)$ is related to a $b_\omega sin(\omega t)$ by Ampere's law in which $b_\omega$ is applied microwave field excitation. In low frequency regime ($\omega<<\omega_p$), the microwave resistance, $R_\omega$ and reactance $X_\omega$ are calculated from the relation [10]:

$$Z_\omega = R_\omega + jX_\omega = \left(\frac{\omega}{2\pi i_{rf}}\right)\int_0^{2\pi/\omega}\int_0^L \dot{\varphi}(x,t)e^{j\omega t}dxdt \qquad (3)$$

Fig. 2 gives application of a dc magnetic field (±*B*) and microwave field $b_\omega sin(\omega t)$ on a Josephson fluxonic diode either symmetrically on both sides of the junction (Fig. 2.a) or asymmetrically (Fig. 2.b). These two cases are the main mechanisms for description of rf-derived JFD which will be compared in detail.

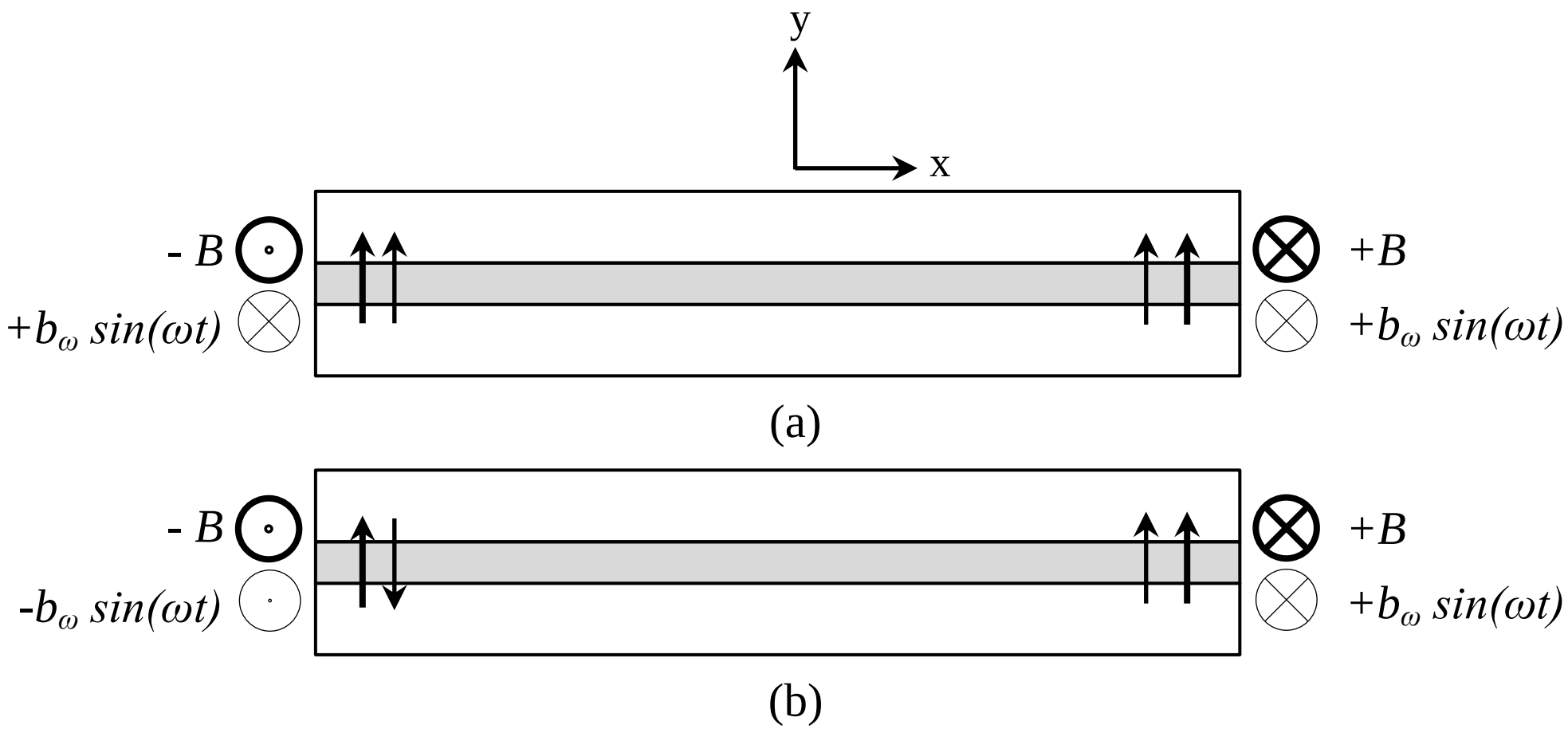

Fig. 2. Boundary condition of JFD with application of spatial reversing dc magnetic field (±*B*) and, (a) symmetric microwave excitation and (b) asymmetric microwave excitation

The detailed study of JFD depends on the parameter values and the boundary conditions depicted in Fig. 2. The above boundary conditions can consider in extended RCSJ model (ERCSJ) to obtain $R_\omega$ and $X_\omega$ by extracting Thevenin's equivalent circuits. ERCSJ model and its equivalent Thevenin circuit are shown in Fig. 3 which connects several Josephson junction in parallel while the same surface impedance of superconductor's plate accounts in series. The surface impedance of a superconductor plate expresses by a complex number whose imaginary part is positive. The real part of this impedance represents the electric loss which manifests itself as a resistance $R_c$ and its positive imaginary part shows that superconductor plate is inductive by L. In this model an ideal Josephson junction behaves like a nonlinear inductor which can be formulated as:

$$L_J(x,t) = \frac{\Phi_0}{2\pi I_c \cos(\varphi(x,t))} \qquad (4)$$

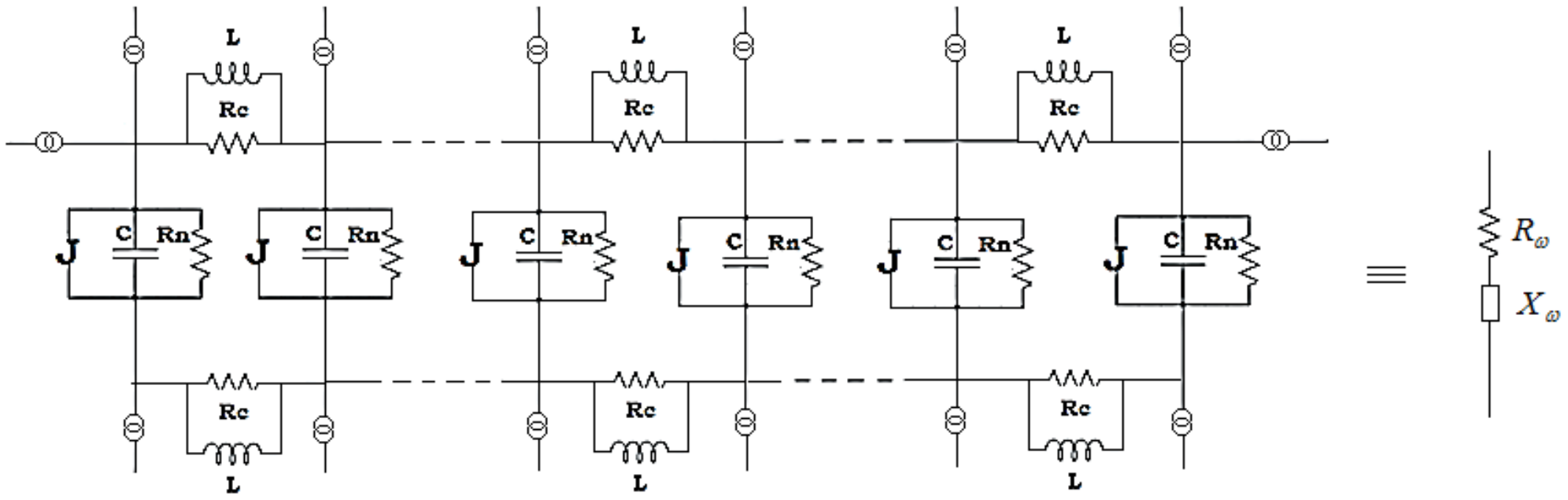

Fig. 3. Extended RCSJ model (ERCSJ) and its equivalent Thevenin circuit

Hence, applying boundary conditions on the first and last cell of ERCSJ model can lead the circuit parameters associated with JFD and find $Z_\omega$ using Thevenin's equivalent circuit. Also, the conditions in Eq. 7, 8 are demonstrated by application of a spatially control current to the central cells of the ERCSJ and removing it from both sides and at a same time a uniform distributed current is added to each cell as bias current.

## III. SIMULATION OF IMPEDANCE

We simulate time evolution of a JFD dynamics with boundary conditions defined in Fig. 2. The algorithm includes totally 1000 Josephson junction cells for $L$=70$\lambda_J$ ($\lambda_J$ is Josephson penetration depth), $J_c$=700 A/cm$^2$ and $\omega_p$ =140 GHz at voltages below the energy gap. For this case, the transition region is defined to be 20$\lambda_J$. In all cases the steady-state behavior is characterized by applying dc control and bias currents to JFD model initially, then waits for steady-state response and in continues rf excitation will apply. So we inspect the solution over at least 10 rf periods, then record $\varphi(x,t)$ and find corresponding inductance for every single time step. Finally, we define the equivalent Thevenin circuit and the impedance is calculated by time-averaging over the last rf cycle. The following sections dedicated to explain the results of each excitation from dc magnetics field, bias current, amplitude and frequency of microwave excitation individually in *A*, *B*, *C* and *D*.

### *A. Control current effect*

In the limits of Thevenin's theorem, JFD acts as an impedance ($\bar{Z}_\omega$=$\bar{R}_\omega$+$j\bar{X}_\omega$) at the presence of both dc and rf drives. The bar above each letter indicates that it is normalized by a factor. For microwave impedance this factor is zero magnetic field microwave impedance $Z_0$=$R_0$+$jX_0$. The first result in Figure 4 shows $\bar{R}_{\omega}$ and $\bar{X}_{\omega}$ versus normalized control current $\bar{I}_{ctrl}$ for different damping parameter 10, 100 and 1000. At this state, no bias is used and $I_{rf}$ sets to be very low avoiding microwave VAV generation with $\omega$=0.1 so no difference between symmetric and asymmetric excitation are pronounced. $\bar{I}_{ctrl}$ is also normalized by $I_{cT}$ and represent the origin of applied dc magnetic field. The level of damping in a Josephson junction is usually determined by $\beta$ which is defines as $\beta=2\pi\rho_i\, t_i J_c/\Phi_0$ where $t_i$ is the junction thickness; $\rho_i$ and $\varepsilon$ are resistivity and permittivity of the junction material, respectively. Large $\beta$ means low damping.

As can be seen from the Fig. 4.a the resistance shows several characteristic features: first of all a threshold field and a subsequent exponential rise above this threshold, with oscillation-like structures. The threshold where a sudden increase in $\bar{R}_{\omega}$ starts is defined as the generation (injection or coupling) of the first VAV at the boundaries. The reactance directly shows dynamic phase slip occurring as the control current increases (Fig. 4.b). These steps are synchronous with adding each new VAV pair by control current and $2\pi$ phase slips meets at boundaries. More clearly, inset plot of Fig 4.b displays the total number of generated VAV pairs at which $2\pi$ phase slip occur. So the value of control current can determine the number of this pairs in JFD. Since coupling of initial VAV pair depends only on the fabrication parameters and is independent of $\beta$ [13] (point 'A' in Fig 4.b inset), generation of extra VAV pair is damping related and for larger $\beta$ the number of produced pairs jumps rapidly which can lead to an increase in $\bar{R}_{\omega}$ (point 'B' in Fig 4.a). Besides, for $\beta$=10, the amplitude of $\bar{R}_{\omega}$ is low and in contrast, once $\beta$ becomes larger, this amplitude increases.

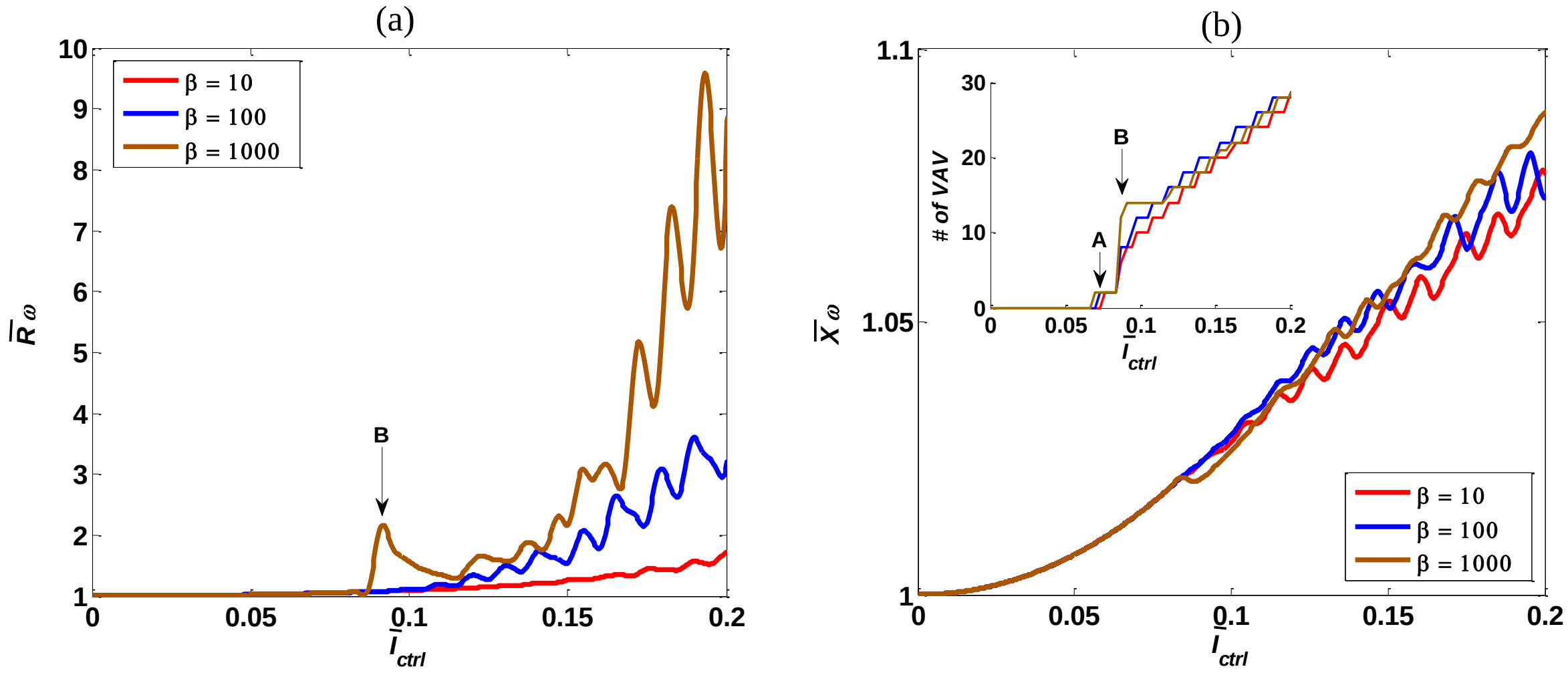

Fig. 4. $\bar{R}_{\omega}$ (a) and $\bar{X}_{\omega}$ (b) versus $\bar{I}_{ctrl}$ for a JFD when $\beta$ =10, 100 and 1000 with very low rf excitation and no bias. The Inset shows the number of generated VAV pairs by increasing $\bar{I}_{ctrl}$. Also, arrows A and B indicate steps at values of $\bar{I}_{ctrl}$ for which the first and more VAV pair generated, respectively.

*B. Bias current effect*

Application of a bias current in JFD sets solitons and anti-solitons in motion by a Lorentz-type force and pushes or pulls them based on bias polarity. At forward bias, depending on $\beta$ they pass through each other or annihilate in transition region. In low damping ($\beta$ =1000), they move and pass through each other while in a JFD with $\beta$ =10 the annihilating will happen [14]. In reverse mode, this condition changes to reflection or elimination of VAV pairs at boundaries, respectively. Under this circumstance, Figure 5 depicts $\bar{Z}_{\omega}$ versus normalized dc bias current ($\bar{I}_{bias}$) for a JFD runs under $\bar{I}_{ctrl}$ =0.15 with very low rf excitation. $\bar{I}_{bias}$ is also normalized by $I_{cT}$. Specified $\bar{I}_{ctrl}$ couples 20 VAV pairs primarily which distribute uniformly along JFD and can move by means of bias current. As can be seen in Fig 5, arrows A, B indicate the steps in figure at which a new VAV generates in addition to primary 20 VAV pairs in forward or reverse bias, respectively. A few important features can be mentioned here. There is a breakdown in reverse bias (point 'A') which is decreases as $\beta$ increase. But in forward, a small sub-threshold exist (point 'B') which is almost independent of β. Fig 5.b describes that when damping parameter is small, breakdown contains a remarkable sharp phase slip in imaginary part of impedance where the number of VAVs increases abruptly after gradual drop of these pairs. This is beginning of some impedance oscillations. These oscillations are periodic but depend on $\beta$ and reflection factor from boundaries their periodicity will affected.

In general results of Fig. 5, increasing $\bar{I}_{bias}$ saturates the number of VAVs in forward bias and this situation persevere for different $\beta$. Remarkably, $\bar{R}_{\omega}$ of forward bias is almost greater than reverse bias in magnitude and damping limits the amplitude of $\bar{R}_{\omega}$ to 34, 206, 645 at breakdown and 50, 208, 1500 for forward sub-threshold respectively for $\beta$ =10, 100 and 1000.

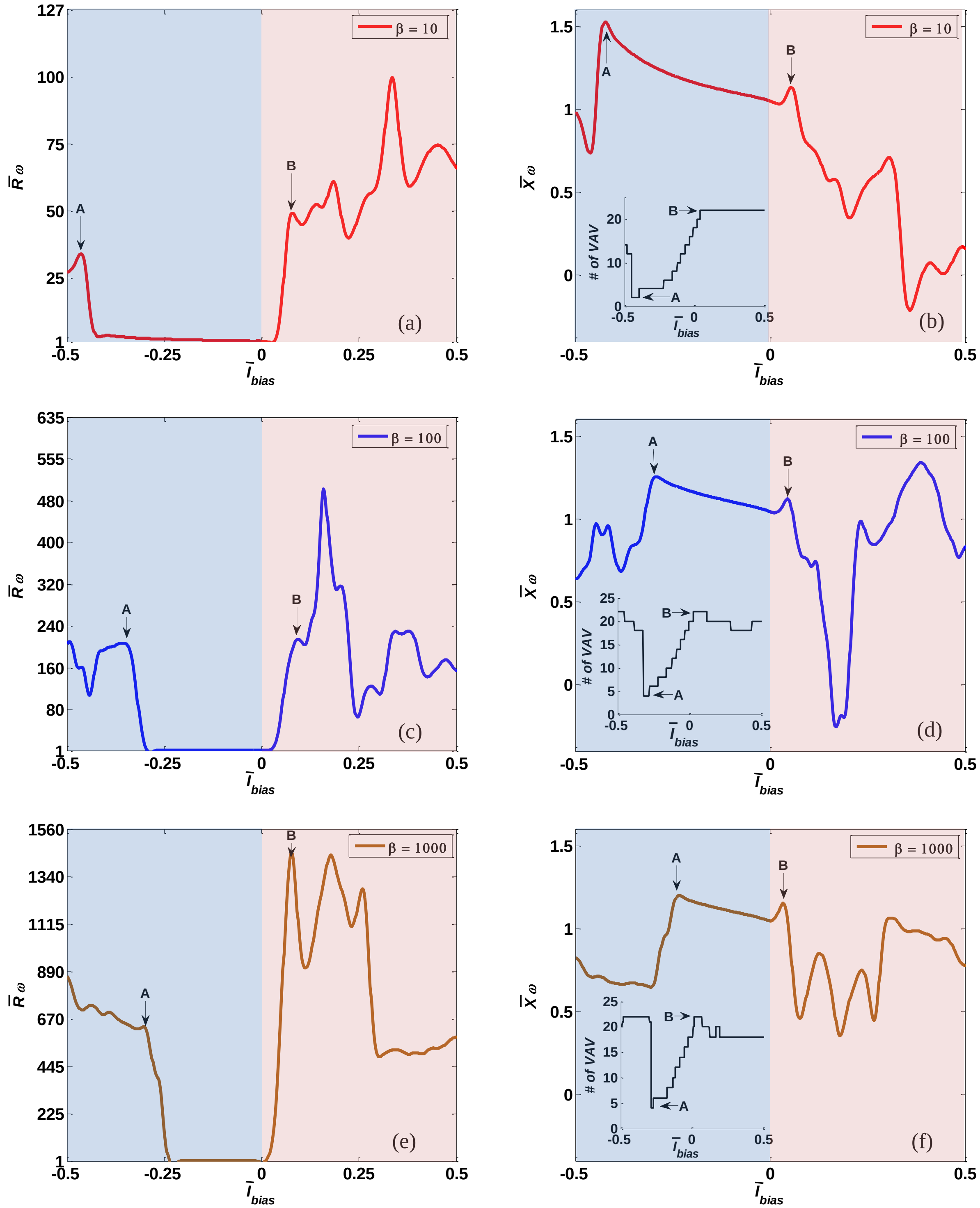

Fig. 5. Real and image part of $\bar{Z}_{\omega}$ versus $\bar{I}_{bias}$ at the presence of dc control current $\bar{I}_{ctrl}$ =0.15 with a very low rf signal for $\beta$ =10 (a, b), 100 (c, d) and 1000 (e, f). The Inset shows the number of generated VAV pairs by increasing $\bar{I}_{bias}$.

### *C. Microwave amplitude effect*

Another important class of investigation is varying amplitude rf excitation ($\bar{I}_{rf}$) while both dc magnetic field and biasing are applied. In this state, microwave field can interact with moving VAVs and even may changes the system harmonics to chaotic in some cases [5]. For this reason, we consider below the case which JFD is biased under specific current and dc magnetic field. As previously discussed, when microwave field is applied on either side of JFD (Fig. 2a), symmetric excitation on boundaries is considered. If the direction of the induced microwave signal is opposite on both sides (Fig. 2b), this situation makes an asymmetric rf field-driven. In Fig. 6, variation of $\bar{R}_{\omega}$ to $\bar{I}_{rf}$ is depicted for these two excitation modes. In this case, Josephson fluxonic diode is reversed biased closely before breakdown (point A in Fig 5) to observe a notable change in $\bar{R}_{\omega}$. As can be seen in Fig. 6, there is a starting point in $\bar{I}_{rf}$ in which the amplitude of microwave excitation can influence on $\bar{R}_{\omega}$ responses. This threshold exists for different values of $\beta$ and depends on how far a JFD biased from point A in Fig. 5. Once $\beta$ is small, this threshold is due to synchronous coupling of the oscillating VAVs with the rf excitation to JFD. If sum of this rf component and the dc bias current exceeds this threshold, first VAV pair is created at the middle of JFD in transition region which result a quick change in impedance. In symmetric drives, it produces equal type vortex at boundaries, but in asymmetric mode, one vortex and one anti-vortex are coupled which are the same as $\bar{I}_{ctrl}$ generated vortex and anti-vortex pair. Under this circumstance, the reverse bias current forces these new carriers to move easily.

However, in symmetric case the oscillating vortexes interact with control current generated vortexes and annihilate each other therefore no effect in steady state response of $\bar{R}_{\omega}$ is obvious (Fig. 6.c). Normally, the value of $\bar{R}_{\omega}$ in an asymmetric mode is higher than symmetric ones. Fluctuations in these figures occur due to the changes in the number of vortices in the JFD and their height is dependent to the microwave drive amplitude $\bar{I}_{rf}$ and $\beta$. For higher values of $\beta$ the steps become closer and are aperiodic (Fig. 6.b, c). The origin of this disorder starts from path of soliton and anti-soliton throw each other. Also, very interesting result is shown in Fig. 6.c for $\beta$=1000. Microwave resistance is about 3 orders of magnitude higher than zero field microwave impedance of a JFD at VAV pair generation edge. Essentially, one can consider that when the $\bar{I}_{rf}$ is high, $\bar{R}_{\omega}$ at specified dc field is high and its value rises when $\beta$ increases. Regardless of low damping JFD, asymmetric rf excitation leads to jump $\bar{R}_{\omega}$ more than symmetric ones at threshold of $\bar{I}_{rf}$.

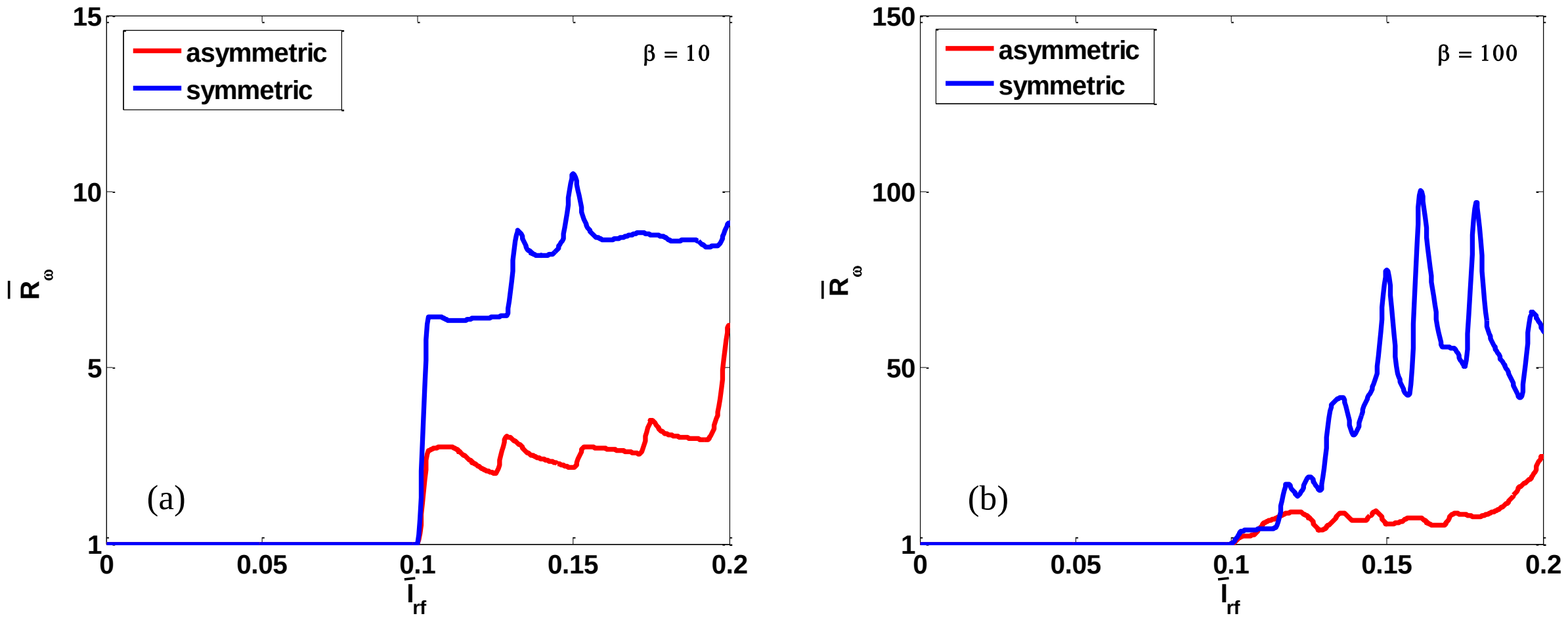

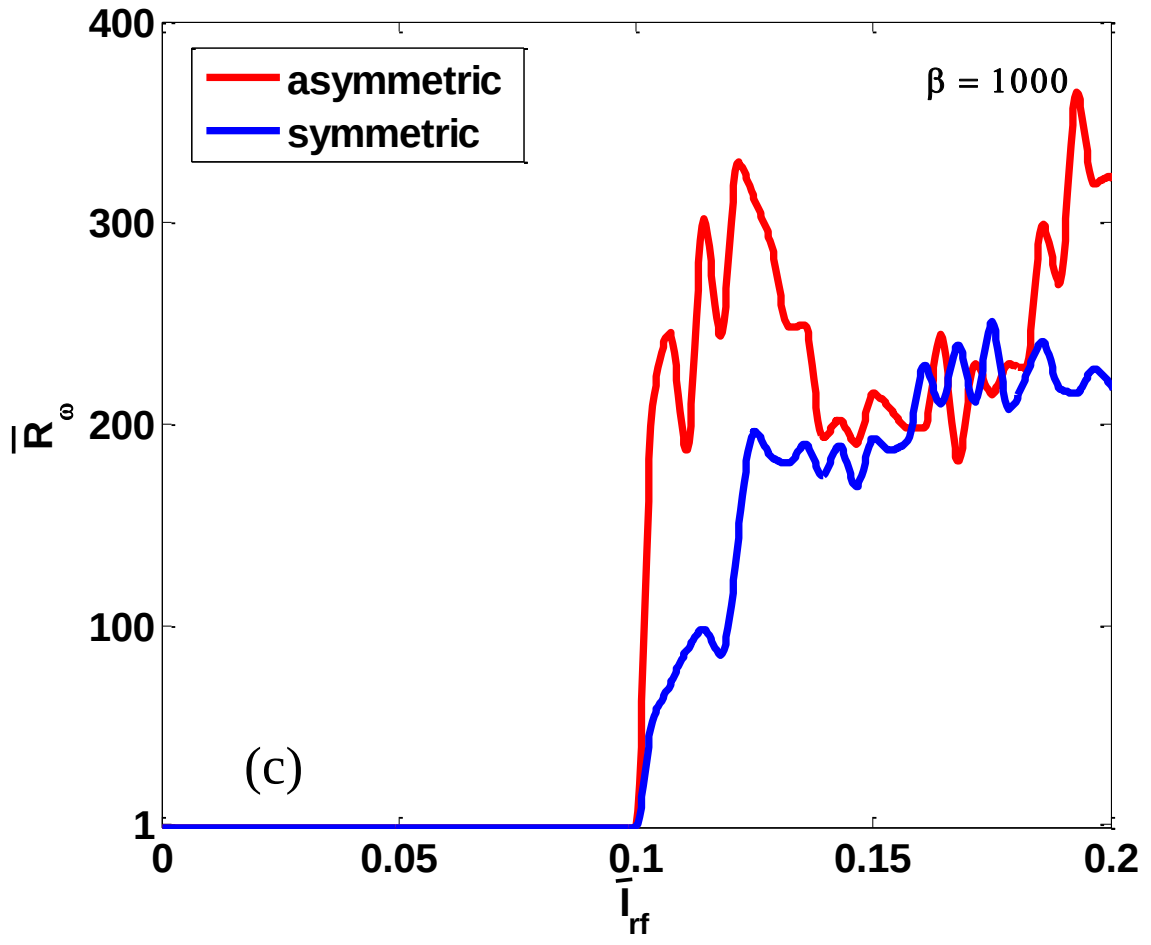

Fig. 6. $\bar{R}_{\omega}$ versus $\bar{I}_{rf}$ at the presence of dc magnetic field ($\bar{I}_{ctrl}$ =0.15) and bias current closely before breakdown in reverse bias (point A in Fig 5) for: $\beta$=10 (a), 100 (b) and 1000 (c). Normally, the value of $\bar{R}_{\omega}$ in an asymmetric mode is higher than symmetric ones and at VAV pair generation edge, is high about three orders of magnitude in comparison with zero field microwave impedance.

### *D. Microwave frequency effect*

The frequency of rf excitation is supplementary factor which influence on microwave impedance directly. This frequency determines the oscillation of coupled solitons to JFD and the result differs from overdamped junction to underdamped ones [15]. Fig. 7 illustrates this dependence. The range of frequency is plotted from 0.1$\omega_p$ to 2$\omega_p$ in a biased JFD with $\bar{I}_{ctrl}$ =0.15 and $\bar{I}_{rf}$ =0.1 in breakdown point. The calculations are, however here, done only for symmetric excitation. It manifests that there is a ramp increase up to plasma frequency for different $\beta$. This value than becomes constant about 17 for $\beta$=10 and variable around 17 for higher $\beta$. Upon overall comparing Fig. 6 with Fig. 7, we see that the amplitude of rf excitation can greatly impact on $\bar{R}_{\omega}$ (up to 300) but its frequency limits this impact to almost 17$R_0$ at frequency higher than plasma frequency.

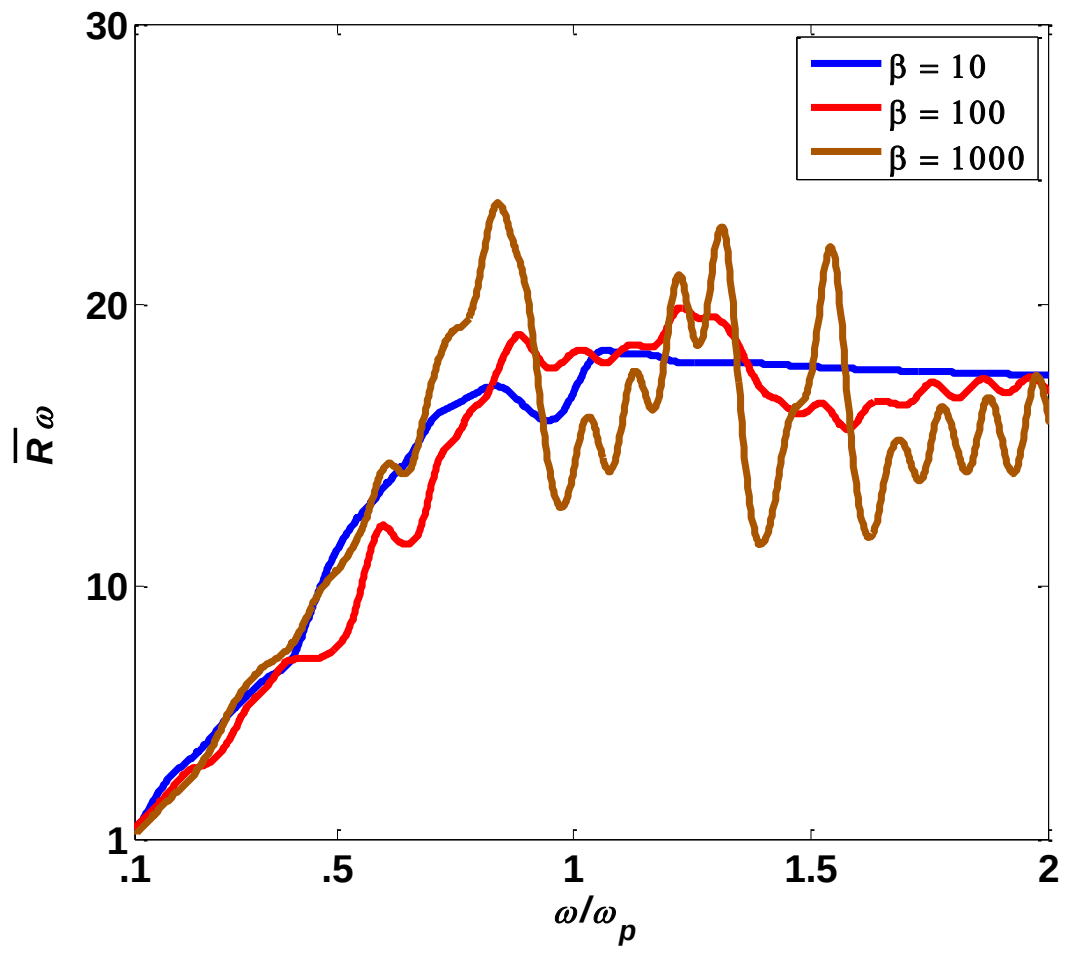

Fig. 7. $\bar{R}_\omega$ versus $\omega/\omega_p$ at the presence of dc magnetic field ($\bar{I}_{ctrl}$ =0.15) and $\bar{I}_{rf}$ =0.1 in breakdown point for symmetric rf excitation.

## IV. CONCLUSIONS

In summary, we have presented the microwave resistance and reactance analysis for Josephson fluxonic diode under various conditions by using ERCSJ model. Based on the results, for a JFD which is excited below its plasma frequency by a very low microwave drive, application of a dc magnetic field increases the number of soliton and anti-soliton pair generation with oscillation-like variation in microwave resistance. These oscillations are due to VAV pair generation and consequent phase slips in total phase difference across the JFD. When $\beta$ is small, there are very sharp and high amplitude variations in impedance of reverse biased JFD at breakdown point and for higher $\beta$, these fluctuations in microwave resistance disappear; however in this case the resistance can increases up to 1500 times more than zero field microwave resistance. Furthermore, the value of $\bar{R}_\omega$ in asymmetric mode is higher than symmetric for low damping parameter and large enough $\bar{I}_{rf}$ and at frequencies below plasma frequency, microwave resistance rises up to a nearly constant value.